\documentstyle[mprocl]{article}

\def\Journal#1#2#3#4{{#1} {\bf #2}, #3 (#4)}

\def\PLB{{\em Phys. Lett.}  B}

\def\PRD{{\em Phys. Rev.} D}

\def\mco{\multicolumn}

\def\ra{\rightarrow}

\def\ko{K^0}

\def\be{\begin{equation}}
\def\ee{\end{equation}}
\def\bea{\begin{eqnarray}}
\def\eea{\end{eqnarray}}

\begin{document}

\title{ACCELERATION-INDUCED CARRIER 
\marginpar{\emph{Based on the first of
two talks given June 23-24, 1997, at the 8th Marcel Grossmann Meeting,
held in Jerusalem, Israel.}}
OF THE IMPRINTS OF GRAVITATION}

\author{ ULRICH H. GERLACH }

\address{Department of Mathematics, Ohio State University, Columbus,
OH 43210, USA}

\maketitle\abstracts{
We exhibit a purely quantum mechanical carrier of the imprints of
gravitation by identifying for a relativistic charge a property which
(i) is independent of its mass and (ii) expresses the Poincare
invariance of spacetime in the absence of gravitation.
This carrier is
a Klein-Gordon-equation-determined vector field given by the
``Planckian power'' and the ``r.m.s. thermal fluctuation'' spectra.
}


Does there exist a purely quantum mechanical carrier of the imprints
of gravitation? 
The motivation for considering this question arises from the following
historical scenario: Suppose the time is 1907 when Einstein had the 
``happiest thought of his life'', which launched him on the path
toward his formulation of gravitation (general relativity).
But suppose Einstein already knew relativistic quantum mechanics, and
that, in fact, he accepted and appreciated it without any reservations
before he started on his journey. How different would his theory
of gravitation have been from what we have today? Put differently,
how different would the course of history have been if Einstein
had grafted relativistic quantum mechanics onto the {\it roots} of
gravitation instead of its trunk or branches?

Nontrivial relativistic quantum mechanics starts with the Klein-Gordon
equation
\begin{equation}
{\partial^2 \psi \over \partial t^2}-
{\partial^2 \psi \over \partial z^2}+ k^2\psi =0,
\label{eq:murnf}
\end{equation}
where $k^2=k^2_x +k^2_y +m^2 $.

The objective of this brief report is to deduce from this equation a
carrier of the imprints of gravitation with the following three
fundamental requirements:

1. The imprints must be carried by the evolving dynamics of a quantum
mechanical wavefunction.

2. Even though the dynamical system is characterized by its
particle mass $m$, the carrier and imprints must \emph{not} depend on the
particle species, i.e. the carrier must be \emph{independent} of $k^2$. This
requirement is analogous to the classical one in which the world line
of a particle is independent of its mass.

3. In the absence of gravitation the carrier should yield measurable
results (expectation values) which are invariant under Lorentz boosts
and spacetime translations.

In quantum mechanics the wave function plays the role which in
Newtonian mechanics is played by a particle trajectory or in
relativistic mechanics by a particle world line. That the wave
function should also assume the task of carrying the imprints of
gravitation is, therefore, a reasonable requirement.

Because of the Dicke-Eotvos experiment, the motion of bodies in a
gravitational field is independent of the composition of these
bodies. Consequently, the motion of free particles in spacetime traces
out particle histories whose details depend only on the gravitational
environment of these particles, not on their internal constitution.
The superposition of different wave functions (states) of a relativistic
particle yields interference fringes which do depend on the mass of a
particle.  If the task of these wave functions is to serve as carriers
of the imprints of gravitation, then, unlike in classical mechanics,
these interfering wave functions would do a poor job at their task:
They would respond to the presence (or absence) of gravitation in a
way which depends on the details of the internal composition (mass) of
a particle. This would violate the simplicity implied by the
Dicke-Eotvos experiment. Thus we shall not consider such
carriers. This eliminates any quantum mechanical framework based on
energy and momentum eigenfunctions because the dispersion relation,
$E^2=m^2 +p^2_z +p^2_y +p^2_x$, of these waves depends on the internal
mass $m$.

Recall that momentum and energy are constants of motion which imply the existence
of a locally inertial reference frame. Consequently, requirement
2. rules out inertial frames as a viable spacetime framework to
accomodate any quantum mechanical carrier of the imprints of
gravitation.  Requirement 2. also rules out a proposal to use
the interference fringes of the gravitational Bohm-Aharanov effect to
carry the imprints of gravitation [1]. This is because the fringe spacing
depends on the rest mass of the quantum mechanical particle.

Requirement 3. expresses the fact that the quantum mechanical carrier
must remain unchanged under the symmetry transformations which characterize
a two-dimensinal spacetime. By overtly suppressing the remaining two spatial
dimensions we are ignoring the requisite rotational symmetry. Steps towards
remedying this neglect have been taken elsewhere [2].

We shall now exhibit a carrier which fulfills the three fundamental
requirements. That carrier resides in the space of Klein-Gordon
solutions whose spacetime domain is that of a {\it pair} of frames
accelerating into opposite directions (``Rindler frames'').  These
frames partition spacetime into a pair of isometric and achronally
related Rindler Sectors $I$ and $II$,
\begin{equation}
\left. \begin{array}{c}
t-t_0 =\pm\xi\sinh \tau \\
z-z_0  = \pm\xi  \cosh \tau
\end{array} \right\} \quad 
\begin{array}{l}
+:~~\hbox{``Rindler Sector I''} \\
-:~~\hbox{``Rindler Sector II''}
\end{array}
\end{equation}
Suppose we represent an arbitrary solution to the K-G equation in the form
of a complex two-component vector normal mode expansion
\begin{equation} \!\!\!\!\!
\left( \!\!\!\!
\begin{array}{c}
\psi_I(\tau,\xi) \\
\psi_{II}(\tau,\xi)
\end{array} 
\!\!\!\! \right)
\!=\!\! \!\int ^\infty _{-\infty} \!\!\!\!\! \!\!\{ a_\omega \!
\left( \!\!\!\!
\begin{array}{c}
1 \\ 0 \end{array} \!\!\! \right)
+b^*_\omega \!
\left( \!\!\!
\begin{array}{c}
0 \\ 1 \end{array} \!\!\! \right)
 \}\sqrt{2\vert \sinh \pi \omega \vert}
{K_{i\omega}(k\xi)\over\pi}e^{-i\omega\tau} d\omega 
\equiv \!\!\int ^\infty _{-\infty} \!\!\!\!\!\!\!\psi_\omega d\omega
\end{equation}
This is a \emph{correlated} (``entangled'') state with two degrees of
freedom. Besides the continuum of boost energies, this state has a
\emph{ discrete polarization} degree of freedom. Its two components
refer to the wave amplitude at diametrically opposite events on a
Cauchy hypersurface $\tau=constant$ in Rindler $I$ and $II$
respectively.  

This representation puts us at an important
mathematical juncture: We shall forego the usual picture of viewing
this solution as an element of Hilbert space with the usual
Klein-Gordon inner product. Instead, we shall adopt a much more
powerful viewpoint based on the vector bundle $C^2\times R$.  Here
$C^2$ is the complex vector space of two-spinors, which is the fiber
over the one-dimensional base manifold 
$R=~\{ \omega:~-\infty < \omega < \infty \} $,
the real line of Rindler frequencies in the mode integral, Eq.(3).

We know that one can add vectors in the \emph{same} vector (fiber)
space.  However, one may not, in general, add vectors belonging to
different vector spaces at different $\omega$'s. The exception is when
vectors in different vector spaces are \emph{parallel}. In that case
one may add these vectors.  The superposition of modes, Eq.(3),
demands that one do precisely that in order to obtain the two
respective total amplitudes of Eq.(3). 

The mode representation of Eq.(3) determines two parallel spinor
fields over $R$, one corresponding to ``spin up'', the other to ``spin
down''. It is not difficult to verify that these two spinor fields are
(Klein-Gordon) orthonormal in each fiber over $R$. The spinor field
\begin{equation}
\{ \left(
\begin{array}{c}
a_\omega \\
b_\omega^*
\end{array}\right) :
~-\infty < \omega < \infty \} \quad .
\end{equation}
is a section of the fiber bundle $C^2\times R$ and it represents a
linear combination of the two parallel vector fields. It is clear that
there is a \emph{one-to-one correspondence between $\Gamma(C^2\times R)$,
the $\infty$-dimensional space of sections of this spinor bundle, and
the space of solutions to the Klein-Gordon equation.} Our proposal is
to have each spinor field serve as a carrier of the imprints of
gravitation: A gravitational disturbance confined to, say, Rindler
$I$ or $II$ would leave its imprint on a spinor field at
$\tau=-\infty$ by changing it into another spinor field at
$\tau=+\infty$.

We know that in the absence of gravitation each of the positive and
negative Minkowski plane wave solutions evolves independently of all
the others.  This scenario does not change under Lorentz boosts and
spacetime translations.  Will the proposed carriers comply with this
invariance, which is stipulated by fundamental requirement 3.?  To
find out, consider a typical plane wave, which in the spinor
representation (3) is a state with a high degree of correlation
between the boost energy and the polarization (``spin'') degrees of
freedom. Suppose for each boost energy we determine the normalized
Stokes parameters of this polarization, i.e. the three Klein-Gordon
based expectation
values of the ``spin'' operator $\overrightarrow \sigma /2$. This is a
three-dimensional vector field over the base manifold $R$, and is given by
[2]
\[
\frac{\langle\psi_\omega,{\overrightarrow \sigma}  \psi _{\omega '}
\rangle}{\langle\psi_\omega,\psi_{\omega '}\rangle}
=\pm \left( \sqrt{N(N+1)},0, \frac{1}{2}+N \right);~~~
N=(e^{2\pi \omega} -1)^{-1};~~-\infty<\omega < \infty
\]
In compliance with requirements 2. and 3., this vector field is (a)
\emph{independent of the particle mass} and (b) the same for \emph{all}
positive (negative) Minkowski plane wave modes, a fact which expresses
its Poincare invariance. The presence of \emph{gravitation would leave
its imprints by producing characteristic alterations in this vector
field.}

Its obvious but noteworthy feature
is that its components coincide with the ``Planckian power'' and the
``r.m.s.  thermal fluctuation'' spectra, in spite of the fact that we
are only considering the quantum mechanics of a single charge.
\par
\noindent \textbf{References}
\par
[1] J.S.Anandan in B.L.Hu, M.P.Ryan, and C.V.Vishveshwara (eds.), {\it
Directions in General Relativity, Volume 1}, (Cambridge University Press,
1993) p.10
\par
[2] U.H.Gerlach in R.T.Jantzen and G.M.Keiser (eds.), {\it The Seventh
Marcel Grossmann Meeting, Part B}, World Scientific Publishing Co. (1996),
{\it ibid} International Jour. of Mod. Phys. 11, 3667 (1996) p.957

\end{document}